\documentclass[aps,11pt,amsfonts,amssymb,amsmath]{revtex4}
\usepackage{enumerate}

\newcommand{\mat}[1]{\underline{\underline{#1}}}
\newcommand{\vek}[1]{\underline{#1}}

\begin{document}
\title{No-go theorem for bimetric gravity with positive and negative mass}

\author{Manuel Hohmann}
\email{manuel.hohmann@desy.de}
\affiliation{Zentrum f\"ur Mathematische Physik und II. Institut f\"ur Theoretische Physik, Universit\"at Hamburg, Luruper Chaussee 149, 22761 Hamburg, Germany}
\author{Mattias N.\,R. Wohlfarth}
\email{mattias.wohlfarth@desy.de}
\affiliation{Zentrum f\"ur Mathematische Physik und II. Institut f\"ur Theoretische Physik, Universit\"at Hamburg, Luruper Chaussee 149, 22761 Hamburg, Germany}

\begin{abstract}
We argue that the most conservative geometric extension of Einstein gravity describing both positive and negative mass sources and observers is bimetric gravity and contains two copies of standard model matter which interact only gravitationally. Matter fields related to one of the metrics then appear dark from the point of view of an observer defined by the other metric, and so may provide a potential explanation for the dark universe. In this framework we consider the most general form of linearized field equations compatible with physically and mathematically well-motivated assumptions. Using gauge-invariant linear perturbation theory, we prove a no-go theorem ruling out all bimetric gravity theories that, in the Newtonian limit, lead to precisely opposite forces on positive and negative test masses.
\end{abstract}
\maketitle

\section{Motivation}
In Newtonian mechanics and gravity the notion of mass appears as a generic term in quite different physical contexts. Taking a closer look one needs to distinguish three different types of mass~\cite{Bondi:1957zz}: active gravitational mass \(m_a\) is the source of gravitational fields; passive gravitational mass \(m_p\) determines the force acting on a test particle in a gravitational field; inertial mass \(m_i\) serves as the proportionality factor relating force and acceleration.

Experiment, however, shows that these at first unrelated types of mass are closely tied together. Both ratios $m_a/m_p$ and $m_i/m_p$ appear to be constant independent of material, see e.g.~\cite{Kreuzer:1968zz,Bartlett:1986zz,Will:1992} and \cite{Eotvos:1922,Roll:1964rd,Nordtvedt:1968qr,Williams:2005rv}. These observations are nicely explained by well-known theoretical principles. Newton's third law asserts that for every force there exists a reciprocal force of equal strength and opposite direction. Considering the gravitational force, this implies that the ratio $m_a/m_p$ between active and passive gravitational mass must be equal for all bodies. The weak equivalence principle states that the acceleration of a physical body in a gravitational field is independent of its composition. This implies that the ratio $m_i/m_p$ between inertial and passive gravitational mass is fixed. It is conventional to choose unit ratios so that all three masses become equal. Taking into account the observational evidence that gravity is always attractive, all mass can be chosen positive.

One may now argue that the experiments and observations mentioned above approve the proportionality and positivity of the different types of mass only for visible matter, i.e., for matter observed through non-gravitational interactions, say through emitted light or other types of radiation. However, assuming Einstein gravity, modern astronomical observations~\cite{Komatsu:2008hk} suggest that visible matter only constitutes a small fraction of about 5\% of the total matter content of the universe. The main constituents of the universe, known as dark matter and dark energy, have not been observed directly. So it remains unknown whether the same fixed ratio relations between the different types of mass are obeyed also in the dark universe. This lack of knowledge hence invites the interesting possibility to modify Newton's third law, the weak equivalence principle, or the positivity of mass.

In this article we will accept this invitation, and consider gravity theories with a modified weak equivalence principle. We will investigate a different ratio between inertial and passive gravitational mass in the dark sector, namely $m_i/m_p=-1$. Assuming that the inertial mass is still positive, we thus have $m_p=m_a$ and $|m_p|=m_i$ in both sectors. This modification introduces negative gravitational masses in such a way that like masses attract while unlike masses repel each other. Applied to cosmology, such a framework could be particularly rewarding. Effects usually attributed to dark matter and dark energy in principle could be explained by the presence of negative masses in the universe. Indeed, if the intergalactic space contained negative mass distributions, these could push positive matter back towards the respective centers of visible galaxies, mimicking the presence of dark matter sources within the galaxies themselves. Moreover, the repulsive gravitational force acting on the visible positive mass galaxies could also lead to accelerating expansion.

Neither the concept of negative mass nor its cosmological motivation are new in the literature, and have been discussed within several theoretical contexts and with different ratios and signs of $m_a$, $m_p$ and $m_i$. Already in 1897, F\"oppl~\cite{Foeppl:1897} introduces negative mass as a natural extension of Newtonian gravity. Consequences within modified Newtonian dynamics are discussed e.g. in~\cite{Milgrom:1986hz}. It has been observed that additional negative masses with \(m_a = m_p = m_i < 0\) neither violate Newton's third law, nor the weak equivalence principle. Forces on such bodies cause these to accelerate in the opposite direction. Probably the most striking example for this behaviour is the gravitational dipole: a system of two bodies of positive and negative mass must forever accelerate in a common direction, the negative mass following the positive one. This effect also exists within Einstein gravity~\cite{Bondi:1957zz,Gibbons:1974zd}, where the ratio \(m_i / m_p = 1\) is manifestly fixed by the weak equivalence principle. Considering only gravitational forces, such that the acceleration of a body is determined purely by the ratio \(m_i / m_p\), the individual signs of \(m_p\) and \(m_i\) do not affect the trajectories. Thus the only possibility to introduce negative mass into Einstein gravity is to choose negative sources for the gravitational field, i.e., \(m_a < 0\). Various properties of such negative mass solutions have been discussed; for instance, gravitational lensing~\cite{Torres:1998cu}, gravitational collapse~\cite{Mann:1997jb} and the stability~\cite{Gibbons:2004au,Gleiser:2006yz} of negative mass black holes. More general repulsive gravitational effects are analyzed e.g. in~\cite{Harari:1990cz,Mannheim:1999bu,Preti:2008zz}.

A consequence of the weak equivalence principle is that all test masses, and all observers, follow the same set of preferred curves, namely geodesics. In other words, there is only one type of observers. Since we wish to relax this condition by allowing the ratio \(m_i / m_p=-1\) for a second type of matter, we have to extend the framework of Einstein gravity. Indeed, it seems natural to introduce a second metric to generate another set of geodesics describing the response of the negative mass observers to the gravitational field. Only then can the gravitational force of a given source be attractive for one class of test particles, while being repulsive for a second class. Such bimetric theories with an antisymmetry between the forces acting on positive and negative masses have become popular under the name `antigravity'~\cite{Hossenfelder:2005gu}, but their consistency, in particular their diffeomorphism invariance, has been doubted~\cite{Noldus:2006jv}. Moreover, the findings of this article place further strong constraints on the construction of such theories: 

We will prove a theorem which rules out the possibility of gravitational forces of equal strength and opposite direction acting on the two classes of test particles in bimetric gravity theories.

The outline of this article is as follows. In section~\ref{sec:theorem}, we will state the assumptions that restrict the class of bimetric gravity theories under consideration. Each of these assumptions is motivated and its physical relevance and implications are discussed. At the end of this section we will formulate the no-go theorem, which will be proven in the following section~\ref{sec:proof}. The proof is based on the most general form of linearized gravitational field equations consistent with our assumptions. Using gauge-invariant perturbation techniques, we will decompose the metric perturbations and identify the components relevant in the Newtonian limit. Then we will show by contradiction that it is not possible to obtain a theory with precisely opposite forces acting on positive and negative matter. We will conclude with a discussion in section~\ref{sec:conclusion}.

\section{Formulation of the theorem and its assumptions}\label{sec:theorem}
We will begin by explaining the theoretical context in which our no-go theorem for `antigravity' is formulated. In particular we will give strong motivation for why we consider bimetric theories in order to describe both attractive and repulsive gravitational forces. Then we will discuss in some detail the assumptions entering the no-go theorem which is formulated at the end of this section. The proof of the theorem will be contained in the following section~\ref{sec:proof}.

We wish to consider gravity theories allowing for positive and negative gravitational masses, so that like or unlike masses attract or repel each other, respectively. Observations tell us that the standard model contains only one type of mass, say the positive type, and we will denote the fields of the standard model collectively by $\Psi^+$. We thus need to introduce a copy $\Psi^-$ of the standard model fields but with negative mass.

Observers follow the curves of massive objects in the small mass limit where back-reaction on the geometric background can be neglected. If fields $\Psi^\pm$ of positive and negative gravitational mass are available, this limit will produce two distinguished classes of curves $\gamma^\pm$ on the spacetime manifold. In extension of standard general relativity, it is reasonable to assume that these curves are described by the autoparallels of two linear connections $\nabla^\pm$.
In order to make measurements, observers attach frames $\{e^\pm_\mu\}$ with $e^\pm_0=\dot \gamma^\pm$ to their spacetime paths $\gamma^\pm$; these frames must be orthonormalized with respect to some metrics $g^\pm$, i.e., $g^\pm(e^\pm_\mu,e^\pm_\nu)=\eta_{\mu\nu}$. In standard fashion, we define observers as inertial, when their frames are non-accelerated and non-rotating; this is equivalent to Fermi-Walker transport according to $\nabla^\pm_{e^\pm_0} e^\pm_\mu=0$. From the orthonormality relation it now follows that the two metrics are covariantly constant, $\nabla^\pm g^\pm=0$. Assuming vanishing torsion of both connections, we thus find that $\nabla^\pm$ are the Levi-Civita connections related to the metrics $g^\pm$.

Note that observers by this construction follow the timelike geodesics defined by just one of the metrics $g^\pm$. To achieve consistency of this notion of causality with that following from the matter field equations we are led to assume that the fields $\Psi^+$ and $\Psi^-$ couple exclusively to the respective metrics $g^+$ and $g^-$. We also assume that there is no non-gravitational interaction between the two types of matter fields, which is consistent with the lack of direct non-gravitational observational evidence for a second type of matter. In other words, matter of type \(\Psi^-\) should appear to be dark from the viewpoint of an observer measuring his world with the metric $g^+$, and vice versa.

This is arguably the most conservative framework for gravity and matter that we can choose to model attractive and repulsive gravitational forces. We simply double the standard ingredients $\Psi^+,\,g^+$ in the standard model by introducing additional negative mass fields $\Psi^-$ and negative mass observers defined with respect to a second metric $g^-$. This yields a theory with two sectors in both of which gravity, in the absence of sources and observers of the second, non-standard type, appears exclusively attractive. The interesting possible implications of the existence of both positive and negative masses are discussed in the introduction, and would only arise from the gravitational interplay of the visible and the dark sector.

We will now formulate, and discuss the physical relevance of, a number of technical assumptions restricting the class of bimetric theories introduced above. These assumptions will be the basic ingredients for our no-go theorem below. For convenience we use in the following underlined quantities to denote two-component vectors (with $+$ and $-$ components), and doubly underlined quantities to denote two by two matrices. Our assumptions are:

\begin{enumerate}[\it (i)]
\item \label{ass:tensor} {\it The gravitational field equations are a set of two symmetric two-tensor equations of the form}
\begin{equation}\label{fieldeqs}
\vek{K}_{ab}[g^+,g^-] = \vek{M}_{ab}[g^+,g^-,\Psi^+,\Psi^-]\,.
\end{equation}

This assumption is consistent with the naive counting argument by which the number of equations agrees with the number of algebraic components of the two metrics $g^\pm$. More importantly, one can think of these gravitational field equations as arising from a combined diffeomorphism invariant matter and gravity action of the form $S_G[g^+,g^-]+S_M[g^+,g^-,\Psi^+,\Psi^-]$. Then variation with respect to the metric $g^+$ would provide the first vector component equation, variation with respect to the metric~$g^-$ the second.

\item \label{ass:derivative} {\it The gravitational tensor $\vek{K}_{ab}$ in the field equations~\emph{(\ref{fieldeqs})} contains at most second derivatives of the metrics $g^+$ and $g^-$.}

This assumption is one of mathematical simplicity. It makes available a number of theorems on the solvability of partial differential equations, as are also relevant in Einstein gravity. Important for us in the proof of our no-go theorem will be the consequent restriction of the number of terms that can appear in the gravitational field equations.

\item \label{ass:sources} {\it The matter source in the field equations~\emph{(\ref{fieldeqs})} is of the form $\vek{M}_{ab}=\mat{J}\cdot\vek{T}_{ab}$, where $\mat{J}$ is a constant invertible matrix, and the components of $\vek{T}_{ab}$ are the respective standard energy momentum tensors $T^\pm_{ab}[g^\pm,\Psi^\pm]$ of positive and negative mass fields.}

This seemingly complicated assumption is easily motivated by recalling that matter fields~$\Psi^+$ should only couple to the metric $g^+$, and fields $\Psi^-$ only to $g^-$. If such field equations come from an action by variation, then this matter action would take the form $S_M[g^+,g^-,\Psi^+,\Psi^-] = j^+ S[g^+,\Psi^+] + j^- S[g^-,\Psi^-]$ for constant $j^\pm\neq 0$. Variation with respect to $g^+$ and $g^-$ then produces precisely the assumed matter source $\vek{M}_{ab}$, the matrix $\mat{J}$ having $j^+,\,j^-$ on the diagonal.

\item \label{ass:flat} {\it The vacuum solution is given by two flat metrics $g^\pm_{ab}=\lambda^\pm \eta_{ab}$ with constants $\lambda^\pm>0$.}

This is another assumption of mathematical simplicity. This vacuum solution has the maximal number of Killing symmetries for both metrics $g^\pm$ simultaneously. The constants $\lambda^\pm$ correspond to the freedom of global rescalings of the Cartesian coordinates. Cosmological constants are excluded; after all, one of the motivations for this framework with both positive and negative mass is the potential explanation of cosmological constants.

\item \label{ass:newton} {\it Stationary solutions with $\partial_0g^\pm_{ab}=0$ exist for arbitrary non-moving dust matter $T^\pm_{00}=\lambda^\pm\rho^\pm$ with small energy densities $\rho^\pm\sim\mathcal{O}(h)$, so that the Newtonian potentials $\phi^\pm$ are small of the same order $\mathcal{O}(h)$ and the (gauge-fixed) linearly perturbed vacuum metrics are}
\begin{equation}\label{ass5}
g^\pm = \lambda^\pm\left[-(1+2\phi^\pm)\,dt\otimes dt + (1-2\gamma^\pm \phi^\pm)\delta_{\alpha\beta}dx^\alpha\otimes dx^\beta\right].
\end{equation}

This simply states that the theory has a (post-)Newtonian limit. Dust matter, non-moving in a given coordinate system, has the energy momentum tensors $T^{\pm\,ab}=\rho^\pm u^{\pm\,a}u^{\pm\,b}$, where $u^\pm\sim\partial_0$. The normalizations $g^\pm(u^\pm,u^\pm)=-1$, corresponding to each type of observer, explain the occurrence of the factors $\lambda^\pm$ in $T^\pm_{00}$. That it should be possible to choose arbitrary small dust distributions $\rho^\pm$ reflects that metric solutions should exist for all appropriate choices of boundary conditions. We shall see below that it is not possible to gauge-fix both metrics $g^\pm$ at the same time to have the form displayed above; the reason is that there are just the standard diffeomorphisms of the manifold available, but there is a second metric tensor. However, there exists a gauge-invariant vector of Newtonian potentials $\vek{I}_1$, and gauges can be chosen so that either $I^+_1=\phi^+$ or $I^-_1=\phi^-$. Note that no specific values for the post-Newtonian parameters $\gamma^\pm$ are assumed, but experiment in our sector of the theory strongly supports the value $\gamma^+=1$~\cite{WillReview}.
\end{enumerate}

These assumptions suitably restrict the class of bimetric gravity theories in which we wish to analyze the behaviour of attractive and repulsive gravitational forces. Using the normalization $8\pi G_N=1$ for Newton's constant, we are now in the position to formulate our no-go theorem.

\vspace{6pt}\noindent\textbf{Theorem.} \textit{We assume a bimetric theory with positive and negative mass sources and observers satisfying points {\it (i)--(v)} detailed above. It is not possible to achieve a Newtonian limit with antisymmetric mass mixing in the Poisson equations for the vector $\vek{I}_1$ of gauge-invariant Newtonian potentials,}
\begin{equation}
\triangle \vek{I}_1 = \frac{1}{2}\Bigg(\begin{matrix}1&-1\\-1&1\end{matrix}\Bigg)\!\!\cdot\!\vek{\rho}\,.
\end{equation}

\vspace{6pt}This is a very surprising no-go statement, and we will prove it in the following section~\ref{sec:proof}. Antisymmetric mass mixing is precisely what one would want from a canonical extension of Newton and Einstein gravity. It implies that the Newtonian force on positive test masses $m$ in any gravitational field is precisely opposite to the force felt by negative test masses $-m$ in the same place. Moreover, positive and negative mass sources generate precisely opposite forces on the same test mass. So the theory excluded by the no-go theorem is exactly that which would allow for a switch of sign in gravitational mass.

\section{Proof of the no-go result}\label{sec:proof}
We will now prove by contradiction the no-go theorem for bimetric gravity theories with positive and negative mass as formulated at the end of the previous section~\ref{sec:theorem}. Since this theorem takes recourse to the Newtonian limit, it is sufficient to use linearized field equations. After discussing the general form of the field equations we will apply the gauge invariant linear perturbation formalism which is known from cosmological perturbation analysis~\cite{Bardeen:1980kt, Malik:2008im,Stewart:1990fm}. This important technique enables us to avoid gauge ambiguities that otherwise might invalidate the proof. We will then show that the scalar, vector, and tensor modes of the metric perturbations decouple. Within the sector of scalar perturbations, which is relevant in the (post-)Newtonian limit, we will finally construct the contradiction constituting the proof.

\subsection{Field equations}
The starting point of our proof is the most general gravitational field equations consistent with the assumptions of the no-go theorem. In agreement with assumptions~{\it (i)}~and~{\it (ii)} these must be symmetric two-tensor equations containing at most second order derivatives of the metric tensors~$g^\pm$. We may easily list all tensorial building blocks that may enter the equations according to their derivative order.
\begin{enumerate}\setcounter{enumi}{-1}
\item No derivatives: the two metrics $g^\pm$, and the endomorphism $j=(g^+)^{-1}g^-$;
\item single derivative: the connection difference $S$ which is a $(1,2)$-tensor field defined by the decomposition $\nabla^-_XY=\nabla^+_XY+S(Y,X)$, so that $S^a_{bc}=\Gamma^-{}^a_{bc}-\Gamma^+{}^a_{bc}$;
\item double derivative: $\nabla^\pm S$, and the two Riemann curvature tensors $R^\pm$ of the two metrics.
\end{enumerate}
Note that terms of the type $\nabla^\pm j$ can be combined from the connection difference $S$ and the metrics; similarly, terms of type $\nabla^\pm\nabla^\pm j$, $\nabla^\pm\nabla^\mp j$ can be combined from derivatives of $S$.

We have already argued that it is sufficient to work with the linearized field equations. Because of assumption~{\it (iv)}, this is a weak field approximation around maximally symmetric Minkowski vacua. Then the metric tensors take the form
\begin{equation}\label{linan}
g^{\pm}_{ab} = \lambda^\pm(\eta_{ab} + h^{\pm}_{ab})
\end{equation}
for constants $\lambda^\pm$. In the course of the following calculation we will keep only terms linear in the perturbations~\(h^{\pm}\), which we assume are of the same order $\mathcal{O}(h)$.

The terms that may now occur in the linearized field equations are the symmetric two-tensors with at most second order derivatives formed from the linearization of the building blocks listed above. Looking at these in more detail one realizes that terms without derivatives cannot appear; that terms of the type $\partial_a h^\pm_{bc}$ always appear quadratic, and so cancel; that all remaining terms are obtained by the various contractions of \(\partial_a\partial_bh^{\pm}_{cd}\). With assumption~{\it (iii)} this leads to the following most general form of the linearized field equations:
\begin{equation}\label{lineq}
\vek{K}_{ab} = \mat{P} \cdot \partial^p\partial_{(a}\vek{h}_{b)p} + \mat{Q} \cdot \square\vek{h}_{ab} + \mat{R} \cdot \partial_a\partial_b\vek{h} + \mat{M} \cdot \partial_p\partial_q\vek{h}^{pq}\eta_{ab} + \mat{N} \cdot \square\vek{h}\eta_{ab} = \mat{J} \cdot \vek{T}_{ab}\,.
\end{equation}
Indices are raised with the metric $\eta$, and $\square=\eta^{pq}\partial_p\partial_q$. The matrices \(\mat{P}\), \(\mat{Q}\), \(\mat{R}\), \(\mat{M}\), \(\mat{N}\) on the geometry side $\vek{K}_{ab}$ of the equations are constant parameters. These are determined by the nonlinear field equations, and also absorb the factors $\lambda^\pm$ in the linearization ansatz~(\ref{linan}). We neither need to know their precise form nor do we need to make any additional assumptions about these matrices to carry out our proof below. 

\subsection{Gauge-invariant formalism}
Since our proof is based on linearized field equations we must take care to ensure that none of our conclusions finally depends on changes of gauge, i.e., on possible changes of coordinates that do not alter the structure of the linearization ansatz~(\ref{linan}) as a small perturbation of maximally symmetric Minkowski vacua. Therefore we will now apply the formalism of gauge-invariant linear perturbation theory known from cosmology~\cite{Bardeen:1980kt, Malik:2008im,Stewart:1990fm} to the ansatz~(\ref{linan}) and equations~(\ref{lineq}).

First, we perform a purely algebraic $(1+3)$-split of the spacetime coordinates $x^a=(x^0,x^\alpha)$ into time and space, and decompose the corresponding components of the perturbations \(\vek{h}_{ab}\) of the metric tensors as
\begin{equation}\label{aldec}
\vek{h}_{00}  =  -2\vek{\phi}\,,\quad
\vek{h}_{0\alpha}  =  \vek{B}_{\alpha}\,,\quad
\vek{h}_{\alpha\beta}  =  -2\vek{\psi}\delta_{\alpha\beta} + 2\vek{E}_{\alpha\beta}\,,
\end{equation}
where \(\vek{E}_{\alpha\beta}\) is trace-free, i.e., \(\delta^{\alpha\beta}\vek{E}_{\alpha\beta} = 0\). Under purely spatial coordinate transformations, the quantities \(\vek{\phi}\) and \(\vek{\psi}\) transform as scalars, \(\vek{B}_\alpha\) as a vector, and \(\vek{E}_{\alpha\beta}\) as a symmetric, trace-free two-tensor. The geometry side $\vek{K}_{ab}$ of the linearized equations~(\ref{lineq}) now decomposes as
\begin{subequations}
\begin{eqnarray}
\vek{K}_{00} &=& 2(\mat{P} + \mat{Q} + \mat{R} + \mat{M} + \mat{N}) \cdot \partial_0^2\vek{\phi} - 2(\mat{Q} + \mat{N}) \cdot \triangle\vek{\phi} - 6(\mat{R} + \mat{N}) \cdot \partial_0^2\vek{\psi}\nonumber\\
&& + 2(\mat{M} + 3\mat{N}) \cdot \triangle\vek{\psi} + (\mat{P} + 2\mat{M}) \cdot \partial_0\partial_{\alpha}\vek{B}^{\alpha} - 2\mat{M} \cdot \partial_{\alpha}\partial_{\beta}\vek{E}^{\alpha\beta}\,,\\
\vek{K}_{0\alpha} &=& (\mat{P} + 2\mat{R}) \cdot \partial_0\partial_{\alpha}\vek{\phi} - (\mat{P} + 6\mat{R}) \cdot \partial_0\partial_{\alpha}\vek{\psi} - \Big(\frac{1}{2}\mat{P} + \mat{Q}\Big) \cdot \partial_0^2\vek{B}_{\alpha}\nonumber\\
&&+ \frac{1}{2}\mat{P} \cdot \partial_{\alpha}\partial_{\beta}\vek{B}^{\beta} + \mat{Q} \cdot \triangle\vek{B}_{\alpha} + \mat{P} \cdot \partial_0\partial^{\beta}\vek{E}_{\alpha\beta}\,,\\
\vek{K}_{\alpha\beta} &=& -2(\mat{M} + \mat{N}) \cdot \partial_0^2\vek{\phi}\delta_{\alpha\beta} + 2\mat{N} \cdot \triangle\vek{\phi}\delta_{\alpha\beta} + 2\mat{R} \cdot \partial_{\alpha}\partial_{\beta}\vek{\phi} + 2(\mat{Q} + 3\mat{N}) \cdot \partial_0^2\vek{\psi}\delta_{\alpha\beta}\nonumber\\
&&- 2(\mat{Q} + \mat{M} + 3\mat{N}) \cdot \triangle\vek{\psi}\delta_{\alpha\beta} - 2(\mat{P} + 3\mat{R}) \cdot \partial_{\alpha}\partial_{\beta}\vek{\psi} - \mat{P} \cdot \partial_0\partial_{(\alpha}\vek{B}_{\beta)}\nonumber\\
&&- 2\mat{M} \cdot \partial_0\partial_{\gamma}\vek{B}^{\gamma}\delta_{\alpha\beta} + 2\mat{P} \cdot \partial^{\gamma}\partial_{(\alpha}\vek{E}_{\beta)\gamma} + 2\mat{Q} \cdot \square\vek{E}_{\alpha\beta} + 2\mat{M} \cdot \partial_{\gamma}\partial_{\delta}\vek{E}^{\gamma\delta}\delta_{\alpha\beta}\,,
\end{eqnarray}
\end{subequations}
where spatial indices are raised with the flat spatial metric $\delta$.

In the second step we perform a differential decomposition of the spatial vectors $\vek{B}_\alpha$ and tensors $\vek{E}_{\alpha\beta}$ in~(\ref{aldec}) according to
\begin{equation}\label{didec}
\vek{B}_{\alpha} = \partial_{\alpha}\vek{\tilde{B}} + \vek{\tilde{B}}_{\alpha}\,, \quad \vek{E}_{\alpha\beta} = \triangle_{\alpha\beta}\vek{\tilde{E}} + 2\partial_{(\alpha}\vek{\tilde{E}}_{\beta)} + \vek{\tilde{E}}_{\alpha\beta}\,,
\end{equation}
where $\triangle_{\alpha\beta} = \partial_{\alpha}\partial_{\beta} - \frac{1}{3}\delta_{\alpha\beta}\triangle$ denotes the trace-free second derivative and
\begin{equation}
\partial^{\alpha}\vek{\tilde{B}}_{\alpha} = \partial^{\alpha}\vek{\tilde{E}}_{\alpha} = 0\,,\quad
\partial^{\alpha}\vek{\tilde{E}}_{\alpha\beta} = 0\,,\quad
\delta^{\alpha\beta}\vek{\tilde{E}}_{\alpha\beta} = 0\,.
\end{equation}
This differential decomposition is unique as has been shown in~\cite{York:1974,Stewart:1990fm}. The essential fact entering the uniqueness argument is that the spatial sections in a Minkowski background are of constant curvature.

In consequence of the algebraic and differential decompositions, the perturbations now are summarized by the so-called scalar modes $\vek{\phi}$, $\vek{\psi}$, $\vek{\tilde B}$, $\vek{\tilde E}$, by the divergence-free (or transverse) vector modes $\vek{\tilde B}_\alpha$, $\vek{\tilde E}_\alpha$, and by the transverse trace-free tensor modes $\vek{\tilde E}_{\alpha\beta}$. These enter the geometry side of the linearized equations as follows:
\begin{subequations}
\begin{eqnarray}
\vek{K}_{00} &=& 2(\mat{P} + \mat{Q} + \mat{R} + \mat{M} + \mat{N}) \cdot \partial_0^2\vek{\phi} - 2(\mat{Q} + \mat{N}) \cdot \triangle\vek{\phi} - 6(\mat{R} + \mat{N}) \cdot \partial_0^2\vek{\psi}\nonumber\\
&&+ 2(\mat{M} + 3\mat{N}) \cdot \triangle\vek{\psi} + (\mat{P} + 2\mat{M}) \cdot \partial_0\triangle\vek{\tilde{B}} - \frac{4}{3}\mat{M} \cdot \triangle\triangle\vek{\tilde{E}}\,,\label{eqn:dec:00}\\
\vek{K}_{0\alpha} &=& (\mat{P} + 2\mat{R}) \cdot \partial_0\partial_{\alpha}\vek{\phi} - (\mat{P} + 6\mat{R}) \cdot \partial_0\partial_{\alpha}\vek{\psi} - \Big(\frac{1}{2}\mat{P} + \mat{Q}\Big) \cdot \partial_0^2\partial_{\alpha}\vek{\tilde{B}}+ \Big(\frac{1}{2}\mat{P} + \mat{Q}\Big) \cdot \partial_{\alpha}\triangle\vek{\tilde{B}}\nonumber\\
&& + \frac{2}{3}\mat{P} \cdot \partial_0\partial_{\alpha}\triangle\vek{\tilde{E}}- \Big(\frac{1}{2}\mat{P} + \mat{Q}\Big) \cdot \partial_0^2\vek{\tilde{B}}_{\alpha} + \mat{Q} \cdot \triangle\vek{\tilde{B}}_{\alpha} + \mat{P} \cdot \partial_0\triangle\vek{\tilde{E}}_{\alpha}\,,\label{eqn:dec:0a}\\
\vek{K}_{\alpha\beta} &=& -2(\mat{M} + \mat{N}) \cdot \partial_0^2\vek{\phi}\delta_{\alpha\beta} + 2\mat{N} \cdot \triangle\vek{\phi}\delta_{\alpha\beta} + 2\mat{R} \cdot \partial_{\alpha}\partial_{\beta}\vek{\phi} + 2(\mat{Q} + 3\mat{N}) \cdot \partial_0^2\vek{\psi}\delta_{\alpha\beta}\nonumber\\
&&- 2(\mat{Q} + \mat{M} + 3\mat{N}) \cdot \triangle\vek{\psi}\delta_{\alpha\beta} - 2(\mat{P} + 3\mat{R}) \cdot \partial_{\alpha}\partial_{\beta}\vek{\psi} - \mat{P} \cdot \partial_0\partial_{\alpha}\partial_{\beta}\vek{\tilde{B}}\nonumber\\
&&- 2\mat{M} \cdot \partial_0\triangle\vek{\tilde{B}}\delta_{\alpha\beta} + \frac{4}{3}\mat{P} \cdot \partial_{\alpha}\partial_{\beta}\triangle\vek{\tilde{E}} + 2\mat{Q} \cdot \triangle_{\alpha\beta}\square\vek{\tilde{E}} + \frac{4}{3}\mat{M} \cdot \triangle\triangle\vek{\tilde{E}}\delta_{\alpha\beta}\nonumber\\
&&- \mat{P} \cdot \partial_0\partial_{(\alpha}\vek{\tilde{B}}_{\beta)} + 2\mat{P} \cdot \triangle\partial_{(\alpha}\vek{\tilde{E}}_{\beta)} + 4\mat{Q} \cdot \square\partial_{(\alpha}\vek{\tilde{E}}_{\beta)} + 2\mat{Q} \cdot \square\vek{\tilde{E}}_{\alpha\beta}\,.\label{eqn:dec:aa}
\end{eqnarray}
\end{subequations}

In the following section~\ref{subsec_dec} we will show that the scalar, vector, and tensor modes in this decomposition completely decouple, i.e., that they lead to equations that can be solved separately. This fact will be important for our proof because it will allow us to set vector and tensor modes to zero. As we will see in the final part of our proof in section~\ref{subsec_cont}, the scalar equations then will provide the crucial information about the (post-)Newtonian limit needed to prove the theorem.

\subsection{Decoupling of modes}\label{subsec_dec}
To demonstrate the decoupling of the scalar, vector, and tensor perturbations we consider in turn the $00$, $0\alpha$ and $\alpha\beta$ components of the linearized equations of motion $\vek{K}_{ab}=\mat{J}\cdot\vek{T}_{ab}$, see~(\ref{lineq}).

A quick inspection of the equation $\vek{K}_{00}=\mat{J}\cdot\vek{T}_{00}$ shows that only scalar modes occur; this becomes obvious from~(\ref{eqn:dec:00}) and by noting that $\vek{T}_{00}$ are scalar modes.

Next consider the equation $\vek{K}_{0\alpha}=\mat{J}\cdot\vek{T}_{0\alpha}=\vek{W}_\alpha$. Clearly the geometry side displayed in~(\ref{eqn:dec:0a}) contains scalar and vector modes; schematically we have
\begin{eqnarray}
\vek{K}_{0\alpha} &=& \partial_\alpha (\textrm{scalar containing only scalar modes})\nonumber\\
&&{}+(\textrm{divergence-free vector containing only vector modes})_\alpha\,.
\end{eqnarray}
Also the matter side $\vek{W}_\alpha$ can be decomposed into appropriate scalar and vector modes, i.e., into a gradient and a transverse vector as $\vek{W}_\alpha=\partial_{\alpha}\vek{\tilde{W}} + \vek{\tilde{W}}_{\alpha}$ with $\partial^{\alpha}\vek{\tilde{W}}_{\alpha} = 0$. The uniqueness of these decompositions on both sides now implies that we obtain two separate equations,
\begin{subequations}
\begin{eqnarray}
\vek{\tilde{W}} &=& (\mat{P} + 2\mat{R}) \cdot \partial_0\vek{\phi} - (\mat{P} + 6\mat{R}) \cdot \partial_0\vek{\psi} - \Big(\frac{1}{2}\mat{P} + \mat{Q}\Big) \cdot \partial_0^2\vek{\tilde{B}}\nonumber\\
&& + \Big(\frac{1}{2}\mat{P} + \mat{Q}\Big) \cdot \triangle\vek{\tilde{B}} + \frac{2}{3}\mat{P} \cdot \partial_0\triangle\vek{\tilde{E}}\,,\label{scalar2}\\
\vek{\tilde{W}}_{\alpha} &=& -\Big(\frac{1}{2}\mat{P} + \mat{Q}\Big) \cdot \partial_0^2\vek{\tilde{B}}_{\alpha} + \mat{Q} \cdot \triangle\vek{\tilde{B}}_{\alpha} + \mat{P} \cdot \partial_0\triangle\vek{\tilde{E}}_{\alpha}\,,\label{vector1}
\end{eqnarray}
\end{subequations}
the first containing only scalar modes, the second only vector modes.

A very similar argument serves to show that the scalar, vector, and tensor modes in $\vek{K}_{\alpha\beta}=\mat{J}\cdot\vek{T}_{\alpha\beta}=\vek{Z}_{\alpha\beta}$ decouple. Both the geometry side explicitly displayed in~(\ref{eqn:dec:aa}) and the matter contribution have to be decomposed as
\begin{equation}
\frac{1}{3}\vek{K}\delta_{\alpha\beta} + \triangle_{\alpha\beta}\vek{\tilde{K}} + 2\partial_{(\alpha}\vek{\tilde{K}}_{\beta)} + \vek{\tilde{K}}_{\alpha\beta} = \frac{1}{3}\vek{Z}\delta_{\alpha\beta} + \triangle_{\alpha\beta}\vek{\tilde{Z}} + 2\partial_{(\alpha}\vek{\tilde{Z}}_{\beta)} + \vek{\tilde{Z}}_{\alpha\beta}
\end{equation}
into scalar modes $\vek{K}$, $\vek{Z}$ determining the traces, further scalar modes $\vek{\tilde K}$, $\vek{\tilde Z}$, transverse vector modes $\vek{\tilde{K}}_{\alpha}$, $\vek{\tilde{Z}}_{\alpha}$, and transverse trace-free tensor modes $\vek{\tilde{K}}_{\alpha\beta}$, $\vek{\tilde{Z}}_{\alpha\beta}$. The important point to observe is that the respective modes on the curvature side only contain contributions from the same type of mode, e.g., the vector $\vek{\tilde{K}}_{\alpha}$ is fully determined by vector modes. The uniqueness of the decomposition on both sides finally yields four separate equations, each containing only a single type of perturbation modes, namely
\begin{subequations}
\begin{eqnarray}
\vek{Z} &=&  2(\mat{R} + 3\mat{N}) \cdot \triangle\vek{\phi} - 6(\mat{M} + \mat{N}) \cdot \partial_0^2\vek{\phi} - 2(\mat{P} + 3\mat{Q} + 3\mat{R} + 3\mat{M} + 9\mat{N}) \cdot \triangle\vek{\psi}\nonumber\\
&&+ 6(\mat{Q} + 3\mat{N}) \cdot \partial_0^2\vek{\psi} - (\mat{P} + 6\mat{M}) \cdot \partial_0\triangle\vek{\tilde{B}} + \Big(\frac{4}{3}\mat{P} + 4\mat{M}\Big) \cdot \triangle\triangle\vek{\tilde{E}}\,,\label{scalar3}\\
\vek{\tilde{Z}} &=& 2\mat{R} \cdot \vek{\phi} - 2(\mat{P} + 3\mat{R}) \cdot \vek{\psi} - \mat{P} \cdot \partial_0\vek{\tilde{B}} + \frac{4}{3}\mat{P} \cdot \triangle\vek{\tilde{E}} + 2\mat{Q} \cdot \square\vek{\tilde{E}}\,,\label{scalar4}\\
\vek{\tilde{Z}}_{\alpha} &=& -\frac{1}{2}\mat{P} \cdot \partial_0\vek{\tilde{B}}_{\alpha} + \mat{P} \cdot \triangle\vek{\tilde{E}}_{\alpha} + 2\mat{Q} \cdot \square\vek{\tilde{E}}_{\alpha}\,,\label{vector2}\\
\vek{\tilde{Z}}_{\alpha\beta} &=& 2\mat{Q} \cdot \square\vek{\tilde{E}}_{\alpha\beta}\,.\label{tensor}
\end{eqnarray}
\end{subequations}

The arguments above show that the decoupling of scalar, vector, and tensor perturbations is essentially a consequence of the uniqueness and cleverness of the algebraic and differential decomposition involved. The relevance of the decoupling will become obvious in section~\ref{subsec_cont}, since it will allow us to limit our discussion to the scalar perturbations which determine the (post-)Newtonian limit. Before we can approach this final part of our proof, however, we need to determine the gauge invariant quantities containing the physical information contained in the metric perturbations.

\subsection{Gauge invariance and consistency}
In this section we will discuss gauge transformations. We will calculate how the scalar, vector, and tensor modes in the metric perturbations change under changes of gauge, and we will find the set of all gauge-invariant quantities. Since gauge transformations are special diffeomorphisms we must require certain consistency conditions so that the gravitational field equations can be rewritten in terms of gauge-invariant quantities only. Otherwise the solutions of the field equations would not be diffeomorphism-invariant.

Gauge transformations in linear perturbation theory are defined as diffeomorphisms that do not change the formal structure of the perturbation ansatz; here this means that the metrics should retain the form of equation~(\ref{linan}),
\begin{equation}
\vek{g}_{ab}=\vek{\lambda}\eta_{ab}+\mathcal{O}(h)\,.
\end{equation}
Every diffeomorphism is generated by a vector field \(\xi\) and changes a tensor field by the Lie derivative; hence $\delta_\xi\vek{g}=\mathcal{L}_\xi\vek{g}$. From the corresponding component expression for the Lie derivative it is clear that the diffeomorphism generated by $\xi$ is a gauge transformation only if $\xi^a\sim\mathcal{O}(h)$. Writing $\xi_a=\eta_{ap}\xi^p$, we then find $\delta_\xi\vek{g}_{ab} = \vek{\lambda} (\partial_a\xi_b+\partial_b\xi_a)$, and so
\begin{equation}\label{gatra}
\delta_\xi\vek{h}_{ab} = \binom{1}{1} (\partial_a\xi_b+\partial_b\xi_a)\,.
\end{equation}

We will now compute how the different components of the metric perturbations \(\vek{h}_{ab}\) transform under such a gauge transformation. As done above for the metric perturbations, we split the gauge transformation $\xi$ into space and time components, and also employ the differential decomposition. We write
\begin{equation}
\xi_0 = \xi, \quad \xi_{\alpha} = \partial_{\alpha}\tilde{\xi} + \tilde{\xi}_{\alpha}
\end{equation}
for a divergence-free spatial vector mode \(\tilde{\xi}_{\alpha}\). According to~(\ref{gatra}), the metric perturbations $h^+_{ab}$ and~$h^-_{ab}$ transform in precisely the same way, namely, as would be the case for a single metric theory. Employing our previous definitions of scalar, vector, and tensor modes we thus obtain for $+$ and~$-$ components the same transformation behaviour under gauge transformations as known from standard calculations~\cite{Stewart:1990fm},
\begin{equation}
\begin{split}\label{gtra}
\delta_\xi\vek{\phi} = -\partial_0\xi&\binom{1}{1}, \quad
\delta_\xi\vek{\psi} = -\frac{1}{3}\triangle\tilde{\xi}\binom{1}{1}, \quad
\delta_\xi\vek{\tilde{B}} = (\partial_0\tilde{\xi} + \xi)\binom{1}{1}, \quad
\delta_\xi\vek{\tilde E} = \tilde{\xi}\binom{1}{1}, \\
&\delta_\xi\vek{\tilde{B}}_{\alpha} = \partial_0\tilde{\xi}_{\alpha}\binom{1}{1}, \quad
\delta_\xi\vek{\tilde E}_{\alpha} = \frac{1}{2}\tilde{\xi}_{\alpha}\binom{1}{1}, \quad
\delta_\xi\vek{\tilde E}_{\alpha\beta} = \vek{0}\,.
\end{split}
\end{equation}

We are now in the position to deduce gauge-invariant linear combinations of modes. A minimal set of such combinations in terms of which all gauge-invariant quantities can be expressed is
\begin{equation}
\begin{split}
\vek{I}_1 = \vek{\phi}+\partial_0\vek{\tilde B}-&\partial_0^2\vek{\tilde E}\,,\qquad
\vek{I}_2 = \vek{\psi}+\frac{1}{3}\triangle\vek{\tilde E}\,,\qquad
I_3 = \tilde{B}^+ - \tilde{B}^-\,,\qquad
I_4 = \tilde{E}^+ - \tilde{E}^-\,,\\
&\vek{I}_\alpha = \vek{\tilde B}_\alpha-2\partial_0\vek{\tilde E}_\alpha\,,\qquad
I'_{\alpha} = \tilde{E}^+_{\alpha} - \tilde{E}^-_{\alpha}\,,\qquad
\vek{\tilde E}_{\alpha\beta}\,.
\end{split}
\end{equation}
Among the gauge-invariant quantities remain six scalars, three vectors, and two tensors. This matches expectations because the gauge transformation, via $\xi,\,\tilde\xi,\,\tilde\xi_\alpha$, contains two scalars and one vector which are eliminated from the originally eight scalars, four vectors, and two tensors in the metric perturbations.

The next step is to find the conditions under which the gravitational field equations can be rewritten in terms of the above gauge-invariants. As discussed previously this is necessary to ensure diffeomorphism-invariance of the solutions. First note that the tensor equations~(\ref{tensor}) are already written in terms of gauge-invariants. We will now illustrate how to proceed for the two vector equations~(\ref{vector1}) and~(\ref{vector2}). We replace all occurrences of $\vek{\tilde B}_\alpha$ by $\vek{I}_\alpha+2\partial_0\vek{\tilde E}_\alpha$, and then $\vek{\tilde E}_\alpha$ by
\begin{equation}\label{subs}
\frac{1}{2}\binom{1}{-1}I'_\alpha + \frac{1}{2}\binom{1}{1}(\tilde E^+_\alpha+\tilde E^-_\alpha)\,.
\end{equation}
The two vector equations now read
\begin{subequations}
\begin{eqnarray}
\!\!\!\!\!\!\!\!\vek{\tilde W}_\alpha & = & -\Big(\frac{1}{2}\mat{P}+\mat{Q}\Big)\!\!\cdot\!\!\left(\partial_0^2\vek{I}_\alpha + \binom{1}{-1}\partial_0\square I'_\alpha\right) +\mat{Q}\cdot\triangle\vek{I}_\alpha + \Big(\frac{1}{2}\mat{P}+\mat{Q}\Big)\!\!\cdot\!\!\binom{1}{1}\partial_0\square (\tilde E^+_\alpha+\tilde E^-_\alpha)\,,\\
\vek{\tilde Z}_\alpha & = & -\frac{1}{2}\mat{P}\cdot\partial_0\vek{I}_\alpha + \Big(\frac{1}{2}\mat{P}+\mat{Q}\Big)\!\!\cdot\!\!\binom{1}{-1}\square I'_\alpha + \Big(\frac{1}{2}\mat{P}+\mat{Q}\Big)\!\!\cdot\!\!\binom{1}{1}\square(\tilde E^+_\alpha+\tilde E^-_\alpha)\,,
\end{eqnarray}
\end{subequations}
and are expressed in terms of gauge-invariant quantities provided that
\begin{equation}\label{matrix1}
(\mat{P}+2\mat{Q})\!\cdot\!\!\binom{1}{1}=\vek{0}\,.
\end{equation}
The procedure of rewriting the scalar equations~(\ref{eqn:dec:00}),~(\ref{scalar2}),~(\ref{scalar3}),~(\ref{scalar4}) is very similar. One expresses $\vek{\phi}$ and $\vek{\psi}$ in terms of $\vek{I}_1$ and $\vek{I}_2$, and then uses $I_3$ and $I_4$ to substitute $\vek{\tilde B}$ and $\vek{\tilde E}$ as in~(\ref{subs}). The additional conditions needed so that the scalar equations only contain gauge-invariants are
\begin{equation}\label{matrix2}
(\mat{P}+2\mat{R})\!\cdot\!\!\binom{1}{1}=\vek{0}\,,\qquad (\mat{M}+\mat{N})\!\cdot\!\!\binom{1}{1}=\vek{0}\,.
\end{equation}
Under these conditions the field equations for the scalar modes become
\begin{subequations}\label{sceqs}
\begin{eqnarray}
\vek{K}_{00} & = & 2(\mat{P} + \mat{Q} + \mat{R} + \mat{M} + \mat{N}) \cdot \partial_0^2\vek{I_1} - 2(\mat{Q} + \mat{N}) \cdot \triangle\vek{I_1} - 6(\mat{R} + \mat{N}) \cdot \partial_0^2\vek{I_2}\nonumber\\
&&+ 2(\mat{M} + 3\mat{N}) \cdot \triangle\vek{I_2} + (\mat{P} + \mat{Q} + \mat{R} + \mat{M} + \mat{N}) \!\cdot \!\!\binom{1}{-1}(-\partial_0^3I_3 + \partial_0^4I_4)\\
&&+ \Big(\frac{1}{2}\mat{P} + \mat{Q} + \mat{M} + \mat{N}\Big) \!\!\cdot\!\! \binom{1}{-1}\partial_0\triangle I_3 - (\mat{Q} - \mat{R}) \!\cdot\!\! \binom{1}{-1}\partial_0^2\triangle I_4 - (\mat{M} + \mat{N})\! \cdot\!\! \binom{1}{-1}\triangle\triangle I_4\,,\nonumber\\
\vek{\tilde{W}} &=& (\mat{P} + 2\mat{R}) \cdot \partial_0\vek{I_1} - (\mat{P} + 6\mat{R}) \cdot \partial_0\vek{I_2} - \Big(\frac{3}{4}\mat{P} + \frac{1}{2}\mat{Q} + \mat{R}\Big) \!\!\cdot\!\! \binom{1}{-1}\partial_0^2I_3\nonumber\\
&&+ \frac{1}{2}\Big(\frac{1}{2}\mat{P} + \mat{Q}\Big) \!\!\cdot\!\! \binom{1}{-1}\triangle I_3 + \Big(\frac{1}{2}\mat{P} + \mat{R}\Big) \!\!\cdot\!\! \binom{1}{-1}\partial_0(\partial_0^2 + \triangle)I_4\,,\\
\vek{Z} &=& -6(\mat{M} + \mat{N}) \cdot \partial_0^2\vek{I_1} + 2(\mat{R} + 3\mat{N}) \cdot \triangle\vek{I_1} + 6(\mat{Q} + 3\mat{N}) \cdot \partial_0^2\vek{I_2} - (\mat{Q} - \mat{R}) \!\cdot\!\! \binom{1}{-1}\partial_0^2\triangle I_4\nonumber\\
&&- 2(\mat{P} + 3\mat{Q} + 3\mat{R} + 3\mat{M} + 9\mat{N}) \cdot \triangle\vek{I_2} + 3(\mat{M} + \mat{N}) \!\cdot\!\! \binom{1}{-1}(\partial_0^3I_3 - \partial_0^4I_4)\\
&&- \Big(\frac{1}{2}\mat{P} + \mat{R} + 3\mat{M} + 3\mat{N}\Big) \!\!\cdot\!\! \binom{1}{-1}\partial_0\triangle I_3 + (\mat{P} + \mat{Q} + \mat{R} + 3\mat{M} + 3\mat{N}) \!\cdot\!\! \binom{1}{-1}\triangle\triangle I_4\,,\nonumber\\
\vek{\tilde{Z}} &=& 2\mat{R} \cdot \vek{I_1} - 2(\mat{P} + 3\mat{R}) \cdot \vek{I_2} - \Big(\frac{1}{2}\mat{P} + \mat{R}\Big) \!\!\cdot\!\! \binom{1}{-1}\partial_0I_3 - (\mat{Q} - \mat{R}) \!\cdot\!\! \binom{1}{-1}\partial_0^2I_4 \nonumber\\
&&+ (\mat{P} + \mat{Q} + \mat{R}) \!\cdot\!\! \binom{1}{-1}\triangle I_4\,.
\end{eqnarray}
\end{subequations}
Now all equations are manifestly rewritten in terms of gauge-invariants only.

We remark that there is a second argument that allows us to understand the matrix conditions~(\ref{matrix1}) and~(\ref{matrix2}) for gauge-invariance: the vacuum equations $\vek{K}_{ab}=\vek{0}$ are tensor equations according to assumption~{\it (i)}, and so should not change under diffeomorphisms, and in particular not under gauge-transformations. Employing the transformation~(\ref{gatra}) in the field equations~(\ref{lineq}) we find the expression
\begin{equation}
\delta_\xi\vek{K}_{ab} = (\mat{P}+2\mat{R})\!\cdot\!\!\binom{1}{1}\partial_a\partial_b\partial^p\xi_p + (\mat{P}+2\mat{Q})\!\cdot\!\!\binom{1}{1} \square\partial_{(a}\xi_{b)} + 2 (\mat{M}+\mat{N})\!\cdot\!\!\binom{1}{1} \square\partial^p\xi_p\eta_{ab}
\end{equation}
which should vanish; the necessary conditions for this precisely agree with~(\ref{matrix1}) and~(\ref{matrix2}).

\subsection{Contradiction}\label{subsec_cont}
We now come to the final part of our proof of the no-go theorem formulated at the end of section~\ref{sec:theorem}. To proceed, we will now employ the remaining assumption~{\it (v)} to simplify the equations derived above. It will be sufficient to consider the scalar perturbations described by equations~(\ref{sceqs}).

Recall that assumption~{\it (v)} says that the theory should have a (post-)Newtonian limit of stationary solutions with respect to the Killing vector field $\partial_0$, and so we may drop all terms containing time derivatives from the equations~(\ref{sceqs}). Moreover, the Newtonian limit should hold for arbitrary non-moving dust sources for which the spatial velocities and internal pressures in the energy momentum tensors vanish; the only non-vanishing components of the energy momentum tensors then are \(\vek{T}_{00} = \mat{\lambda}\cdot\vek{\rho}\), for $\mat{\lambda}=\left(\begin{smallmatrix}\lambda^+&0\\0&\lambda^-\end{smallmatrix}\right)$ and energy densities \(\vek{\rho}\). Also, the metric solutions in suitable gauges should be given by~(\ref{ass5}) for some post-Newtonian parameters $\gamma^\pm$.

The last point implies a very useful relation between gauge-invariants, namely $\vek{I}_2 = \mat{\gamma}\cdot\vek{I}_1$ for $\mat{\gamma}=\left(\begin{smallmatrix}\gamma^+&0\\0&\gamma^-\end{smallmatrix}\right)$. To see why this is true, consider the form of the metric $g^+$ with scalar perturbations, which is
\begin{equation}
g^+ = \lambda^+\left[-(1+2\phi^+)dt\otimes dt+2\partial_\alpha\tilde B^+dt\otimes dx^\alpha+\left((1-2\psi^+)\delta_{\alpha\beta}+2\triangle_{\alpha\beta}\tilde E^+\right)dx^\alpha\otimes dx^\beta\right].
\end{equation}
It is clear from the gauge transformations~(\ref{gtra}) that we can choose $\xi$ and $\tilde\xi$ so that both $\tilde B^+=0$ and $\tilde E^+=0$; this gauge choice is called the longitudinal gauge. The metric $g^+$ then can be compared with the assumed post-Newtonian form
\begin{equation}
\lambda^+\Big[-(1+2\phi^+)dt\otimes dt + (1-2\gamma^+ \phi^+)dx_\alpha\otimes dx^\alpha\Big]
\end{equation}
relevant for non-moving dust~\cite{WillReview}, and linearized in the Newtonian potential $\phi^+$. Therefore, in longitudinal gauge, a given value of $\gamma^+$ for the observer related to $g^+$ implies $\psi^+=\gamma^+\phi^+$. The corresponding gauge-invariant statement is $I_2^+=\gamma^+ I_1^+$. Repeating the same argument for the observer related to~$g^-$ then also shows that $I_2^-=\gamma^- I_1^-$.

The following argument is essentially unchanged by the value of the post-Newtonian parameters~$\gamma^\pm$. In order to simplify the presentation maximally, we choose $\gamma^\pm=1$ in which case $\vek{I}_2 = \vek{I}_1$. Then the equations (\ref{sceqs}) for the scalar perturbations simplify under the assumption~{\it (v)} to
\begin{subequations}
\begin{eqnarray}
\!\!\!\!\!\!\mat{J}\cdot\mat{\lambda}\cdot\vek{\rho} & = & -2(\mat{Q}-\mat{M}-2\mat{N}) \cdot \triangle\vek{I_1} - (\mat{M} + \mat{N}) \!\cdot\!\! \binom{1}{-1}\triangle\triangle I_4\,,\\
 \vek{0}& = &(\mat{P} + 2\mat{Q}) \!\cdot\!\! \binom{1}{-1}\triangle I_3\,,\\
 \vek{0}& = & -2(\mat{P} + 3\mat{Q}+2\mat{R}+3\mat{M}+6\mat{N}) \cdot \triangle\vek{I_1} + (\mat{P} + \mat{Q} + \mat{R} + 3\mat{M} + 3\mat{N})\!\cdot\!\! \binom{1}{-1}\triangle\triangle I_4\,,\\
 \vek{0} & = &-2(\mat{P}+2\mat{R}) \cdot \vek{I_1} + (\mat{P} + \mat{Q} + \mat{R}) \!\cdot\!\! \binom{1}{-1}\triangle I_4\,.
\end{eqnarray}
\end{subequations}
We now eliminate the term containing $\triangle\triangle I_4$ in the third equation by substituting an appropriate combination of the first and last equations. This yields the simple result
\begin{equation}
-4\mat{Q} \cdot \triangle\vek{I_1} = \mat{J}\cdot\mat{\lambda} \cdot \vek{\rho}\,,
\end{equation}
from which our contradiction will now follow. We must consider two possible cases:
\begin{enumerate}
\item
\(\mat{Q}\) is not invertible. In this case, the dimension of the image of \(\mat{Q}\), viewed as an endomorphism of \(\mathbb{R}^2\), is less than two. From assumption~{\it(v)} we know that \(\vek{\rho}\) can be chosen arbitrarily; since \(\mat{J}\) and $\mat{\lambda}$ by assumptions~{\it (iii)} and~{\it (iv)} are invertible, it follows that \(\mat{J}\cdot\mat{\lambda} \cdot \vek{\rho}\) spans $\mathbb{R}^2$. This is a contradiction.

\item
\(\mat{Q}\) is invertible. In this case, we obtain the equation
\begin{equation}\label{poisson}
\triangle\vek{I_1} = -\frac{1}{4}\mat{Q}^{-1}\cdot\mat{J}\cdot\mat{\lambda} \cdot \vek{\rho}\,,
\end{equation}
which is the Poisson equation for the two Newtonian potentials $\vek{I}_1$ of the two different observers related to the metrics $\vek{g}$. Antisymmetric mass mixing, as defined in the no-go theorem, occurs if and only if
\begin{equation}
-\frac{1}{2}\mat{Q}^{-1}\cdot\mat{J}\cdot\mat{\lambda} = \Bigg(\begin{array}{cc}
1 & -1 \\
-1 & 1
\end{array}\Bigg)
\end{equation}
Since the left hand side of the equation is invertible while the right hand side is not, this immediately leads to the desired contradiction.
\end{enumerate}
This concludes the proof of the Theorem of section~\ref{sec:theorem} that the construction of bimetric theories with antisymmetric mass mixing is not possible.~$\square$

\section{Conclusion}\label{sec:conclusion}
In this article, we have argued that the most conservative framework for a theory of `antigravity' containing both positive and negative gravitational mass sources and observers is a bimetric theory with two copies of standard model matter. The second metric generates another set of geodesics describing the response of negative mass observers to the gravitational field. The two matter sectors only interact gravitationally so that the negative mass sector appears dark from the point of view of a positive mass observer. Such theories could offer a potential explanation for the unaccounted dark part of the universe.

As our central result we have proven a no-go theorem ruling out all canonical bimetric extensions of Einstein gravity. In these excluded theories the gravitational forces acting on the two different types of matter are precisely opposite. Moreover, opposite matter sources yield opposite forces on a given test mass. Our proof proceeds from the most general form of the linearized field equations for a bimetric theory. Using gauge-invariant linear perturbation theory, we have computed the (post-)Newtonian limit. We have obtained the Poisson equations for the Newtonian potentials, provided that the sources of the gravitational fields are constituted by non-moving dust matter. It turns out that each copy of standard model matter contributes to both Newtonian potentials, but with an important restriction: the mixing matrix, which governs the contribution of each matter source to each gravitational field, is manifestly invertible. However, opposite forces on the different types of matter require a non-invertible mixing, and so are excluded.

Of course the strength of any no-go-theorem highly depends on its assumptions. Our assumptions as stated in section~\ref{sec:theorem} are strongly motivated on physical and mathematical grounds, and apply to a very large class of bimetric theories. Hence our no-go theorem is highly restrictive and puts very severe constraints on the construction of bimetric `antigravity'. 
Now that we have a clear picture of what is not possible, we might wonder whether `antigravity' theories exist at all. We will now discuss some possibilities using a smaller, or different, set of assumptions so that our no-go theorem does not apply.

One simple way of avoiding the conclusion of our no-go theorem is to allow for different strengths of the gravitational forces acting on positive and negative test masses in the same gravitational field. In the Newtonian limit, the Poisson equation would then read
\begin{equation}
\triangle \vek{I}_1 = \frac{1}{2}\Bigg(\begin{matrix}1&-\alpha\\-\alpha&1\end{matrix}\Bigg)\!\!\cdot\!\vek{\rho}
\end{equation}
for $\alpha\neq \pm 1$. Now the mixing matrix that determines the contribution of the matter sources to the Newtonian potentials is invertible. So the only cases excluded by the proof of our no-go theorem are $\alpha=1$ which corresponds to exactly opposite forces, and $\alpha=-1$ which means equal force on all observers (this is the situation modelled by Einstein gravity with a single metric). One may argue, however, that the introduction of an additional parameter $\alpha$ does not present a canonical extension of Einstein gravity.

A second possibility is to relax the assumption that the sources of the gravitational field originate from the standard action for matter fields. Instead one might use different actions containing both metric tensors, which would change the matter side of the equations. Of course, this would also change the equations of motion for matter fields so that all types of matter would be influenced by both metrics. This is problematic because it would change the causality of field propagation. However, one might argue that our observations of matter in gravitational fields are limited to particular settings, e.g. to the solar system. The theory might be constructed so that the changes in causality there might be weak or even cancel completely. In other words, our assumption~(\ref{ass:sources}), which restricts the matter side of the field equations, could be valid within the bounds of current observations, but may not hold in general.

Third, we may consider a less conservative framework containing more than two metric tensors and a correspondingly higher number of standard model copies. The computation we have performed in our proof of the no-go theorem can be generalized to this case. It turns out that the Poisson equation in the Newtonian limit is formally the same as equation~(\ref{poisson}). As in the bimetric case, we could now demand that like masses attract while unlike masses repel each other with equal strength. This corresponds to the requirement that the Poisson equation should be
\begin{equation}\label{poisson2}
\triangle \vek{I}_1 = \frac{1}{2}\left(\begin{matrix}
1 & -1 & \cdots & -1\\
-1 & 1 & & -1\\
\vdots & & \ddots & \\
-1 & -1 & & 1
\end{matrix}\right)\!\!\cdot\vek{\rho}\,.
\end{equation}
If $n$ metrics and $n$ copies of the standard model are used, the mixing matrix has non-vanishing determinant \((2 - n) \cdot 2^{n - 1}\) for $n\neq 2$. Hence it is invertible, and the conclusion of the no-go theorem only applies to the bimetric case $n=2$.

Finally we wish to mention the recent comment~\cite{Hossenfelder:2009hb} on this article, where it is argued that the `antigravity' theory of~\cite{Hossenfelder:2008bg} also avoids our no-go theorem. This comes from the simple fact that this theory contains two additional (1,1)-tensor fields besides our two metrics and two copies of the standard model.

The above discussion demonstrates that several ways of circumventing our no-go theorem might exist. Further research now must show whether one of these can be realized, i.e., whether it is actually possible to construct a concrete `antigravity' theory with attractive and repulsive gravitational forces. Once such a theory is available, its physical implications will have to be investigated. It would be particularly interesting to study the predictions of the theory for cosmology, and whether the extra copies of the standard model may serve as an explanation for the dark universe.

\acknowledgments
The authors are happy to thank Sabine Hossenfelder, Niklas H\"ubel, Christian Pfeifer, Christian Reichwagen, Felix Tennie and Lars von der Wense for interesting discussions. MH gratefully acknowledges full financial support from the Graduiertenkolleg 602 `Future Developments in Particle Physics'. MNRW gratefully acknowledges full financial support from the German Research Foundation DFG through the Emmy Noether fellowship grant WO 1447/1-1.


\end{document}